GRAPHENE, MECHANICAL AND STRUCTURAL PROPERTIES AND DEVICES, COMPOSITES, COMPUTATIONAL CHEMISTRY

c.galiotis@iceht.forth.gr

# Failure Processes in Embedded Monolayer Graphene under Axial Compression

Charalampos Androulidakis[1,2], Emmanuel N. Koukaras[2], Otakar Frank[3], Georgia Tsoukleri[2,4], Dimitris Sfyris[2], John Parthenios[2,4], Nicola Pugno[5], Konstantinos Papagelis[1,2], Kostya S. Novoselov[6], and Costas Galiotis[1,2,4*]

**Exfoliated monolayer graphene flakes were embedded in a polymer matrix and loaded under axial compression. By monitoring the shifts of the 2D Raman phonons of rectangular flakes of various sizes under load, the critical strain to failure was determined. Prior to loading care was taken for the examined area of the flake to be free of residual stresses. The critical strain values for first failure were found to be independent of flake size at a mean value of –0.60 % corresponding to a yield stress of -6 GPa. By combining Euler mechanics with a Winkler approach, we show that unlike buckling in air, the presence of the polymer constraint results in graphene buckling at a fixed value of strain with an estimated wrinkle wavelength of the order of 1–2 nm. These results were compared with DFT computations performed on analogue coronene/ PMMA oligomers and a reasonable agreement was obtained.**

1. Department of Materials Science, University of Patras, Rio Patras, 26504 (Greece)

2. Institute of Chemical Engineering Sciences, Foundation of Research and Technology-Hellas (FORTH/ICE-HT), Stadiou Street, Platani, Patras Acahaias, 26504 (Greece)

3. J. Heyrovsky Institute of Physical Chemistry, v.v.i., Academy of Sciences of the Czech Republic, 182 23 Prague 8, Czech Republic

4. Inter-departmental Program in Polymer Science & Technology, University of Patras, Rio Patras, 26504 (Greece)

5. Department of Civil, Environmental and Mechanical Engineering, Università di Trento, via Mesiano, 77 I-38123

6. School of Physics and Astronomy, University of Manchester, Manchester, U.K.

Graphene consists of a two-dimensional (2D) sheet of covalently bonded carbon and forms the basis of both 1D carbon nanotubes, 3D graphite but also of important commercial products, such as, polycrystalline carbon (graphite) fibres. As a single defect-free molecule, graphene is predicted to have an intrinsic tensile strength higher than any other known material[1] and tensile stiffness similar to values measured for graphite. Indeed recent experiments[2] have confirmed the extreme stiffness of graphene of 1 TPa and provided an indication of the breaking strength of graphene of 42 N $m^{-1}$ (or 130 GPa assuming graphene thickness of 0.335 nm). These experiments involved the simple bending of a tiny flake by an indenter on an AFM set-up and the force-displacement response was approximated by considering graphene as a clamped circular membrane made by an isotropic material. To date there are no reported data, as yet, on pure axial stretching of graphene monolayers to fracture. Furthermore there is still an uncertainty concerning the ultimate tensile strain to failure which expected to be higher than even 30% making graphene a very ductile material indeed in spite of its very high stiffness.

Previous studies by us and others have reported the effect of applied strain on the Raman spectrum (G and 2D bands) under compression and tension loading histories [3, 4]. For axial tension the brittle beam systems used to impart deformation to the graphene flakes are not suitable for inducing graphene tensile failure (assumed to be as high as 25–30%[2]). Other systems employed such as PDMS elastomers[5], suffer from poor bonding between graphene and matrix and therefore interfacial slippage initiates at relatively low strains and that hampers the strain transfer efficiency. The situation however in axial compression is different as the material itself fails at low strains due to its infinitesimally small thickness. Initial results reported earlier on a limited size range of only 3 graphene flakes[3], indicated that the critical strain to failure depends on the Euler geometric term[6] but at an effective (graphene) bending stiffness that was orders of magnitude higher than the assumed value in air.

In this present study we extend the work reported previously by examining a wide range of graphene sizes (length-to-width ratio) embedded in the SU8/ PMMA matrix system. Special attention is given to the efficient transfer of stress from the polymer to the inclusion (monolayer graphene) as assessed by the value of Raman wavenumber shift per % of applied axial strain. For the interpretation of the acquired data, we consider the general problem of buckling of an embedded plate for which the "bonding" between the flake and the surrounding medium is modelled by linear elastic springs that act only in the z-direction (Winkler's approach). A constitutive equation yielding the critical strain to compression failure as a function of the graphene elastic constants, the dimensions of the graphene flakes and the Winkler's modulus is derived. The same methodology has been used earlier in the modelling of instability problems in the case of embedded carbon nanotubes[7]-[8]. Moreover, the Winkler's approach has also been used to study vibrations[9], buckling[10] and wave propagation[11] of embedded single layer graphene sheets, the vibrational[12] and the wave propagation[13] characteristics of embedded double and multi- layer[14] graphene sheets using nonlocal elasticity theory.

From the full range of experimental results the model predicts a universal strain value for failure, for all reported flakes regardless of their specific dimensions. This is because the critical strain to failure in compression is mainly affected by the presence of polymer and the magnitude of graphene-polymer interaction (interface) which is expressed by the Winkler modulus. Thus, it is concluded that the graphene mode of failure in the embedded state differs dramatically from that observed in the air. Hence, significant differences in the values of the half-waves determined for the two media (air and polymer) are expected in the case of graphene flakes.

Finally, in order to obtain an independent value for the Winkler modulus we have performed calculations of the interaction between monolayer graphene and PMMA within the framework of density functional theory (DFT). These calculations can be separated in two parts, (a) geometry optimisations of isotactic PMMA (i-PMMA) and syndiotactic (s-PMMA) helixes with a length of two rings each, and (b) rigid potential energy surface (PES) scans of various relative configurations of a PMMA monomer and Coronene. The optimised structures are in good agreement with the experimental findings of Kumaki[15] et al. The structural details of the optimized structures were used to determine the PMMA–Coronene relative configurations for the PES scans. The PES scans, the theoretical (optimised) structures and the experimental findings of ref 15 were used to produce the interaction energy curves which we furthermore fitted to a suitably modified Lennard–Jones type potential. With these primitive (intermediate) potentials of the PES scans we have constructed composite potentials, that approximate the theoretically maximum values of the non-linear spring stiffness per unit area, $K$, of PMMA (in its various forms) and graphene.

## Results and Discussion

The graphene samples were prepared by mechanical exfoliation of HOPG using the well-established scotch-tape method [16]. The graphene flakes are embedded in a SU8/ PMMA layer [3]. As mentioned in Methods all specimens were compressed using a four point bending jig (see **Figure 1**a) to ensure a better control of the imparted strain up to values high enough to induce graphene compressive failure. In most experiments Raman measurements were taken mainly from the geometric centre of the flakes (both $x$ and $y$ directions). However, if the dimension of the flake along the loading direction, $x$, is less than that required for efficient stress transfer then the applied strain values given by the beam formula are not necessarily attained at the positions from which Raman measurements are taken. The reason for this is that the stress/strain in an inclusion embedded in a polymer matrix is built through shear at the interface. As found already by a number of authors, in order to transfer efficiently the applied stress or strain from the matrix to graphene a transfer length of the order of 1–2 μm is needed [17]. Indeed if the available length for stress transfer in the axial direction is less than twice the required transfer length (see Supporting Information) then only a fraction of the applied stress/strain is transmitted to the flake. Furthermore the same effect is observed if the laser probe interrogates an area adjacent to the ends of a larger flake from which the applied stress/ strain is built-up (see Supporting Information). It is nevertheless possible to devise a method to convert the nominal applied strains to "actual" graphene strains by comparing the slope of the measured wavenumber shift per strain to the mean universal value, which for the 2D Raman peak has been found to be 60± 5cm$^{-1}$/%, through the formula (see Supporting Information):

$$\varepsilon_{graphene} = \varepsilon_{applied}\left(\frac{\left(\frac{\partial \Delta \nu}{\partial \varepsilon}\right)_{measured}}{\left(\frac{\partial \Delta \nu}{\partial \varepsilon}\right)_{\max imum}}\right)_{T,\ \varepsilon=0} \quad \text{or} \quad \varepsilon_{graphene}(\%) = \frac{\varepsilon_{applied}(\%)}{|60|}\left(\frac{\partial \Delta \nu}{\partial \varepsilon}\right)_{measured,\ \varepsilon=0} \qquad (1)$$

The mean value and corresponding error were calculated from flakes with sufficient stress transfer; i.e. with uncorrected slope values from **Table 1** larger than 55cm$^{-1}$/%. For this correction to be valid care must be taken to collect the Raman data from the same position at each strain level and to conduct the measurements at a constant temperature (preferably RT). Also the above formula is only valid up to the first inflection point as failure processes that are triggered at that point will also affect the shift of the Raman wavenumber.

In **Figure 2**a, b plots of wavenumber versus applied strain are shown for two different flakes, one with a sufficient length to result in efficient stress transfer (length, *l*=30μm) and another (given in **Figure 2**b) with a smaller length (*l*=4μm), which corrections need be applied to. The results in Figure 2a are fitted by a 4$^{th}$ degree polynomial which accurately captures both the slopes at zero and critical strains and the slopes near the origin. The critical strain value for buckling shown at the plateau corresponds to the point of zero slope. For the case of Figure 2a the initial slope $\left(\frac{\partial \Delta \nu}{\partial \varepsilon}\right)_{T,\ \varepsilon=0} = 56.4\ \frac{\text{cm}^{-1}}{\%}$ indicates efficient load transfer which is not surprising since the flake has a length of 30 μm and the data were obtained at the middle of the flake. The position of the 2D peak increases with applied strain until a plateau is observed. After this plateau the 2D peak relaxes in value with any further increment of applied strain. For this flake the strain at which the plateau occurs corresponds to a critical strain of failure of ~−0.7%.

In contrast, the results of Figure 2b must be corrected since the length of the flake is of the order of the critical length (~2 $L_t$) and the initial slope is – as expected-38 cm$^{-1}$/% i.e. outside the bounds of the required value of 60 cm$^{-1}$/%. As seen the corrected data through formula (1) yield a critical value of compressive failure of ~–0.45% for this flake. A full collection of the data obtained in this work and in Ref [3] are summarised in **Table 1**.

A plate with low flexural rigidity fails under compression in air by elastic buckling. When the plate is supported/ embedded, out-of-plane buckling is restricted by the presence of the surrounding material which, in effect, provides a strong support against any buckling instability. The result is that the critical stress for buckling of the plate is orders of magnitude larger than that of the free plate. Analogous phenomena have been observed in embedded microtubules [18],[19]. It is worth mentioning here that in carbon (graphene) fibre reinforced composites, compression failure initiates as a result of internal (shear) collapse of the fibre prior to Euler (instability) buckling. So in other words the elastic instability observed in the case of a graphene monolayer is the upper limit for composite failure. In other words, reinforcing materials that do not suffer from internal collapse perform much better in compression than commercial reinforcements such as carbon or Kevlar fibres.

In **Figure 3** the experimental data of the critical strain to compressive failure determined here but also those presented in an earlier publication are plotted as a function of the aspect ratio of *l/w* of the rectangular flake. As shown the data fluctuate around a mean value of about –0.6% regardless of aspect ratio. With reference to experimental scatter, this increases considerably for values of *l/w* close to the origin (very short flakes) for which certain corrections were made based on the initial slope of the Raman wavenumber vs. strain line. The scatter reflects the difficulty in measuring the graphene strain at that region due to the inefficient stress transfer as mentioned above. However, there is no doubt that the average value even for very short flakes is within the range of the mean value obtained from all the data points of the graph.

A widely used method for modelling plates resting on elastic foundation is that of Winkler's [20] in which the interaction between the plate and the foundation is modelled with discrete linear elastic springs acting only in the z-axis. No interaction between adjacent springs is considered. The reaction pressure of the foundation is linearly related to the deflection of the plate in the vertical direction:

$$p_r = K_w u(x,y) \tag{2}$$

Where $p_r$ is the reaction pressure of the polymer matrix, $K_w$ is the Winkler's foundation modulus in units of stress per unit deflection ($N/m^2/m$), and *u(x,y)* is the deflection in the z- direction. We note that nonlinear frameworks can also be found in the literature for modelling the substrate effect [39,40]. However, for relatively small strains the substrate employed here (glassy polymer) exhibits a linear stress-strain curve and, hence, there is absolutely no need to resort to a non-linear framework. Moreover, the axial stress is imparted to the inclusion through direct compression loading and not via the release of a prestressed substrate as in the case of Refs. [39, 40]).

Here, for the embedded graphene sheets, the flake is assumed to be pushing on both surfaces against linear springs as shown in **Figure 4**. Thus the PMMA is modelled as an elastic medium which is an appropriate assumption since the experiments were performed at small strains where the response of the polymer is elastic. Following the analysis of Timoshenko [6] for the case examined here, one can derive analytically the compressive strain of failure for plates resting on elastic foundations. We start from the energy balance per area, *A*, of the compressed plate which is given by [6]:

$$E = U_b + U_f - T \tag{3}$$

where $E$ is the total energy of the system- assumed zero at the point of failure-$U_b$ is the plate bending energy, $U_f$ is associated with the elastic strain energy as the flake is pushing/ pulling against the surrounding matrix and $T$ is the axial compression energy released by the flake buckling. The corresponding expressions for the terms of equation 3 are given by equation 4-6 (see Supporting Information):

$$U_b = \frac{D}{2}\int_A \left\{ \left( \frac{\partial u^2}{\partial x^2} + \frac{\partial u^2}{\partial y^2} \right)^2 - 2(1-v)\left[ \frac{\partial u^2}{\partial x^2}\frac{\partial u^2}{\partial y^2} - \left( \frac{\partial u^2}{\partial x \partial y} \right) \right] \right\} dA \quad (4)$$

$$U_f = \frac{K_w}{2}\int_A u^2 dA \quad (5)$$

$$T = \frac{1}{2}\int_A N_x \left( \frac{\partial u}{\partial x} \right)^2 dA \quad (6)$$

where $N_x$ is the compressive force per unit length applied in x-direction, $D$ is the flexural rigidity and $v$ is the Poisson's ratio of the plate and $K_w$ is the Winkler modulus. For a simply supported plate, $K_w$ is the spring constant that describes the interaction between the plate and the foundation and $w$ is plate's width. The boundary conditions that should be satisfied are $u(0,y)=u(x,0)=0$, $u'_x(0,y)=u'_y(x,0)=0$. For the out-of-plane displacement $u$ we make the assumption [6,39,40,41,42] that a sinusoidal wave is formed during buckling of an embedded flake with unconstrained ends:

$$u(x,y) = \sum_{m=1}^{\infty}\sum_{n=1}^{\infty} u_{mn} \sin\left( \frac{m\pi x}{l} \right) \sin\left( \frac{n\pi y}{w} \right) \quad (7)$$

Where *m, n* are the half-waves of the buckling mode in the *x* and *y* directions, respectively. Such an assumption for the displacement seems to be reasonable even though we have no direct observation of how the graphene plate is deformed due to the intervening polymer layer. However, useful information can be extracted from simply supported flakes that confirm the sinusoidal nature of graphene buckling under axial compression (**Figure 5**).

By inserting equation 7 into the balance of energy equation 3 and after some further manipulation (see Supporting Information) we arrive at the following constitutive expression for the axial critical strain to failure:

$$\varepsilon_{cr} = \pi^2 \frac{D}{C}\frac{k}{w^2} + \frac{l^2}{\pi^2 C}\left(\frac{K_w}{m^2}\right) \tag{8}$$

In the above relation $C$ is the tension rigidity of the flake which has been found to be 340 $Nm^2$ while the Euler geometric term $k$ is defined as follows

$$k = \left(\frac{mw}{l} + \frac{l}{mw}\right)^2 \tag{9}$$

As it can be seen from equation 8, in the absence of polymer ($K_w=0$) the second term is zero and therefore the problem reduces to the Euler buckling formula for a freely suspended (graphene) plate in air.

The number of half-waves, $m$, is evaluated by equating the force expressions (see SI) for two consecutive buckling modes[6], i.e.

$$N^m = N^{m+1} \tag{10}$$

This renders for the number of half waves:

$$m^2(m+1)^2 = \frac{l^4}{w^4} + \frac{l^4 K_w}{\pi^4 D} \tag{11}$$

It is evident from Equations 8, 9 and 11, that the Winkler's approach requires either the modulus, $K_w$ or the number of half-waves, $m$, in order to yield analytically the critical strain to buckling in compression. All other parameters, such as the elastic constants ($D$ and $C$) [2, 21] and the flake dimensions, are known. As mentioned earlier, the experimentally obtained $\varepsilon_{cr}$ is insensitive to the ratio of $l/w$ and retain a value of ~-0.6% for a wide range of sizes and axial geometries (Figure 3). Hence, it is facile to estimate a value of $K_w$ (and hence $m$ from equation 11) from the measured $\varepsilon_{cr}$ and to compare with values reported in the literature for similar systems. The results are presented in **Table 1** from which a value of Winkler's modulus for the embedded monolayer graphene in the PMMA/ SU8 system of $K_w$=6.7GPa/nm is obtained. This value is of the same order of magnitude with the value of $K_w$= 1.13 GPa/nm for a system polymer-graphene in ref.[11] and a value of 7.2 GPa/nm that corresponds to the stiffness of van der Waals forces between graphene and Si [22].

Another outcome of this work is that our model predicts a high amount of half waves for the buckling mode for the embedded case. Using simple geometrical arguments (see SI) we can evaluate the amplitude and the wavelength of the out of plane displacement. For the amplitude the estimated value is of approximately 0.61Å. Amplitude values in this range have been experimentally observed [23] (~0.5Å) for graphene flakes suspended over trenches on a copper substrate under thermally induced compression. This demonstrates the ability of graphene to exhibit sub-nanometer buckling amplitudes. The buckling half wavelength can be calculated by dividing the final length of the buckled sheets with the number of half-waves:

$$\lambda = \frac{l\left(1-\varepsilon_{cr}\right)}{m} \qquad (12)$$

The results presented in **Table 1** indicate clearly that the wavelength of the embedded monolayer graphene flakes is of the order of 1.2 nm which agrees well with the value of 2.68 nm reported previously [22] for a simply supported graphene under axial compression. We note that this wavelength is essentially the half-wavelength since it is derived from the half-wave number. The full wavelength therefore corresponds to the value of approximately 2.4 nm. In fact, the value obtained here is expected to be smaller than the rippling wavelength of a graphene flake in air as shown schematically in **Figure 6**, since for a constrained (here by polymer matrix) plate to bend, it should push into/ pull apart the surrounding matrix. Hence, from the energetic point of view, short-wavelength buckling will be preferred because the same degree of end-to-end compression is possible with smaller lateral motion (i.e. less energy required for that mode of deformation). Multiple rippling under compression in supported monolayer graphene sheets has been investigated using mixed atomistic- finite element simulations[24]. Analogous phenomena have been observed for a whole variety of specimens such as embedded rods[25], cytoskeletal microtubules[18], living animals such as snakes[26] etc. It is worth noting here that the graphene region under investigation at the middle of the flake is considered to be wrinkle-free at the onset of the experiment since, due to the preparation procedure (spin coating at high speeds), the graphene is stretched in all directions and Raman frequency mapping prior to loading ensures that only regions that exhibit no residual strain are examined. Multiple wrinkling under compression in supported monolayer graphene sheets has been investigated using mixed atomistic-finite element simulations[24].

We have modelled the interaction of the graphene sheet with the polymer matrix by considering two independent terms; one corresponding to the graphene–polymer van der Waals interactions and the other describing the influence of the elastic deformations of the matrix itself at the vicinity of the interface. For each case we have ascribed a spring of certain stiffness. These two springs are considered to be connected in series and with the overall stiffness equal to the reduced stiffness of the two springs (for details see Supporting Information).

In order to get an independent estimate of the strength of the van der Waals interactions between the PMMA polymer and graphene we have performed calculations within the framework of density functional theory (DFT). The primitive potentials (see supporting information figure S4), $u_{1-4}$, were computed from potential energy surface scans of the MMA molecule at various distances from coronene which is a good analogue of graphene for computational purposes. The PES scans were performed on four relative orientations of MMA with respect to coronene, which are the most commonly occurring configurations for depositions of isotactic PMMA (*i*-PMMA) and syndiotactic PMMA (*s*-PMMA) over a graphene substrate. These MMA monomer–coronene relative configurations for the PES scans are shown in figure S3. In order to maintain high-accuracy in the PES scans while improving the computational efficiency the calculations were performed employing the B97-D functional, which includes dispersion corrections, along with a high quality basis set, and the final values were suitably scaled as described in the Methods section. The required structural details were obtained through geometry optimizations of *i*-PMMA and *s*-PMMA with trans-gauche and trans-trans backbone conformations respectively[27]. The geometry optimizations were performed on a $16_l$*i*-PMMA chain (total number of atoms 245 with 132 H-atoms) and a $32_l$*s*-PMMA chain (total number of atoms 485 with 260 H-atoms) using the B97-D functional and employing the def-SVP[28] basis set. The resulting optimized structures, which are helical as expected[29], are shown in **Figure 7**.

All of the potentials, i.e. the initial potentials from the PES scans and the final composite potentials, have been fitted to the same modified Lennard–Jones type potential of the general form:

$$U(z) = 4\varepsilon\left(\left(\frac{\sigma}{z}\right)^{12C} - \left(\frac{\sigma}{z}\right)^{6C}\right) \tag{13}$$

In the case of the primitive potentials, $u_{1-4}$ from the PES scans, the parameter $z$ corresponds to the transverse distance from the coronene plane of the monomer's carbon atom nearest to the plane. In the case of the composite potentials, the function $U$ corresponds to the potential per unit area (Å$^{-2}$).

It is straightforward to find the non-linear spring stiffness per unit area:

$$K(z) = \frac{\partial^2}{\partial z^2}U(z) = 24\frac{\varepsilon C}{\sigma^2}\left(2\left(12C+1\right)\left(\frac{\sigma}{z}\right)^{12C+2} - \left(6C+1\right)\left(\frac{\sigma}{z}\right)^{6C+2}\right) \tag{14}$$

At the equilibrium position, $z_{\text{eq}}$, we have $\sigma/z_{\text{eq}} = (1/2)^{1/6C}$, and

$$K(z_{eq}) = 24\frac{\varepsilon C}{\sigma^2}\left(2\left(12C+1\right)\left(\frac{1}{2}\right)^{(12C+2)/(6C)} - \left(6C+1\right)\left(\frac{1}{2}\right)^{(6C+2)/(6C)}\right) \tag{15}$$

The modified Lennard–Jones potential fitting parameters of the interaction energy curves are given in the supporting information. The primitive potentials are shown in figure S4 (see SI). The final potentials have been constructed as a linear combination of the primitive potentials from the PES scans. The coefficients of the expansion were determined from the structural details of the optimized *i*-PMMA and *s*-PMMA helices (details are given in the SI) and from the AFM images of Kumaki *et al*. This cluster-model based approach of performing calculations on fundamental and compact blocks of the system followed by an empirical synthesis of the interactions corresponding to the extensive system permits the use of, and calibration to, high-accuracy methods such as those described in the Methods section. These composite potentials have been fitted to the modified Lennard–Jones form of equation 13, given per unit area and are shown in figure 8.The fitting parameters and the non-linear spring stiffness per unit area, *K*, for each case are given in **Table 2**. Overall, our ideal model assumes dense deposition, flat adsorption scheme, rigid polymer axis, and perfectly clean surfaces, all of which are factors that may reduce (some may do so significantly) the *K* value of the true (experimental) system.

These values correspond to the case which PMMA interacts with one side of the graphene surface. However, in our mathematical model we have considered interactions on both sides of the graphene. For this reason we have additionally performed a two-parameter PES scan on a coronene with two MMA molecules, one on each of the coronene surfaces. The configuration corresponds to that used in PES1, and the scanning parameters are the distances $R_1$, $R_2$ between either monomer or the coronene plane. The resulting scan is shown in figure S5 of the supporting information. The presence of the second MMA molecule reduces the interaction very slightly, and the overall interaction energy is 1.2 % less compared to twice the interaction energy of the corresponding single monomer case. Taking these factors this into account the theoretical maximum values for *K* are 21.77 and 40.17 GPa/nm, for *s*-PMMA and *i*-PMMA respectively. Our ideal model assumes dense deposition, flat adsorption scheme, rigid polymer axis, and perfectly clean surfaces, all of which are factors that may reduce (some do so significantly) the *K* value of the true (experimental) system. An independent evaluation of the second spring length is non-trivial. Thus, we have based our treatment of this term on experimental results while also taking advantage of our (DFT based) calculation for the first spring. Specifically, we form a family of curves (see Supporting Information) from which estimates for an effective spring length for the second spring can be obtained. This analysis permits for a comprehensive and integrated treatment of quantities from different scales. Overall, the agreement between theory and experiment is satisfactory bearing in mind the assumptions of the DFT calculations and the assumed nature of vdW interactions between graphene and polymer (eg clean surfaces, decrease polymer porosity etc.).

## Conclusions

The compression stability of embedded monolayer graphene flakes under axial compression was examined experimentally by means of Raman spectroscopy and analytically by employing a combined Euler–Winkler approach. A large range of graphene aspect ratios was tested. Care was taken to ensure that the graphene was resting flat on the substrate and no residual strain or Raman frequency fluctuations within the region under investigation were detected at the onset of the experiment. For very short specimens in the loading direction it was important to assess the effect of transfer length upon the efficiency of stress transfer and to apply a transfer length correction to the obtained raw data. Computational DFT methods were also used in order to assess independently the level of interaction between graphene and PMMA. The results have shown clearly that unlike a flake compressed in air, the critical strain to failure for an embedded graphene is not affected by the flake dimensions due to the presence of the polymer constraint. The form of the buckled graphene at the instability cannot be observed but evidence is provided that the flake undergoes a small half wavelength (~2nm) multiple wrinkling (buckling) on the imposition of a compressive load of approximately −0.6%. The agreement between experiment and the DFT work was satisfactory bearing in mind the limitations of the DFT methods.

## Methods

*Sample preparation*. Graphene monolayers were prepared by mechanical cleavage of HOPG (High Order Pyrolitic Graphite) and transferred onto PMMA bars. The PMMA bars have dimensions of 2.9 mm thickness, width of 12 mm, length of 10 or 12 cm and covered on the top by a ~200 nm thick layer of SU8 photoresist (SU8 2000.5, MicroChem). The graphene samples were first located using an optical microscope and monolayers with the desired dimensions were chosen for testing. The number of layers was identified with Raman measurements. Finally a layer of PMMA (495 PMMA A 3, MicroChem) was spin coated on the top with 6000 rpm. It is noted that in the previous work the samples were spin coated with another polymer S1805 photoresist (Shipley). Graphene and the polymer matrix do not chemically interact so the results are not influenced and no change in the Raman measurements was observed. For compression measurements on an embedded graphene it is of paramount importance for the flake or the region investigated to rest absolutely flat on the polymer substrate. Bearing in mind that the surface of polished polymers might exhibit a roughness of at least tens of nanometers, care was taken to map carefully the area under investigation (approx. 10 $\mu m^2$) to ensure that the Raman wavenumber of the examined region corresponds to the stress-free values and no fluctuations of Raman frequency that would possibly indicate graphene wrinkling, were detected.

*Mechanical testing*. The new experimental data reported here were carried out with a four-point-bending machine adjusted to the MicroRaman (InVia Reflex, Renishaw, UK) laser. The reason for switching from cantilever to 4-point-bend testing is to allow a much better control of the strain imparted to graphene regardless of its position on the beam and to thus allow a higher level of applied strain that previously. The measurements were recorded with an excitation wavelength 785 nm and laser power ~1 mW. More details can be found in references 3, 4. The top surface of the bars is compressed by deflecting the beam on the vertical direction. The deflection of the beam is related with the compressive strain by eq. 16:

$$\varepsilon(\delta) = 4.48\delta t / L^2 \tag{16}$$

Where $\varepsilon$ is the strain, $\delta$, $t$are the deflection and the thickness of the beam respectively and $L$ is the length of the supporting span. The samples were placed in the middle of the PMMA bar, thus the maximum compressive strain with a uniform distribution in the middle of the bar is achieved. The strain was applied with an increment step of ~0.05 %. Raman spectra were recorded after every increment of the strain. The values of strain of equation 16 have been confirmed with measurements taken with strain gauge. For flakes that exhibit residual stress (compressional) at the onset of the experiment (zero applied strain), the beam is first flexed in tension so as to reach zero strain in graphene and then the data are collected in compression from that point onwards. This way a complete curve starting from zero graphene strain to first failure and beyond is established. AFM images were recorded using a Dimension Icon Microscope (Bruker) with ScanAsyst Air tips in the PeakForce tapping mode.

*DFT analysis*. All-electron density functional theory calculations including dispersion corrections (DFT-D2) were performed using the generalized gradient functional B97-D[30] of Grimme. Results of the B97-D functional were tested against higher accuracy methods and at different levels of theory. Specifically, we compared results near the equilibrium of the PESs produced by B97-D with those using the Grimme's double-hybrid functional B2PLYP-D[31] as well as with spin-component scaled second-order Møller–Plesset perturbation theory (SCS(MI)-MP2)[32],[33].The quality of the results obtained by SCS(MI)-MP2 for intra- and inter-molecular interaction has been shown for a wide range of systems[33,[34]] to be comparable to methods of much higher computational cost such as coupled–cluster with single and double and perturbative triple excitations (CCSD(T)) for dispersion type interactions. The optimized scaling parameters that we used are $c_{OS}$ = 0.17 and $c_{SS}$ = 1.75[33].The B2PLYP-D results were practically the same with the SCS(MI)-MP2 results as can be seen on figures SI-4a and SI-4b. The results from the B97-D functional were only slightly overestimated compared to the SCS(MI)-MP2 results by about 11%. In all cases the high quality triple-$\zeta$ def2-TZVPP[35] basis set was employed. The empirical parameterization of DFT-D methods partially accounts for basis set superposition errors (BSSE) and counterpoise (CP) corrections[36] [8] are not needed as long as properly polarized triple zeta basis sets are used, such as the ones used here[30,[37],[38]]. However, for the calculations using SCS(MI)-MP2 it is necessary to account for basis set superposition errors (BSSE) and we have done so using the counterpoise correction method[36].

Tight convergence criteria were enforced on the SCF energy ($10^{-7}$ au), the one electron density (rms of the density matrix up to $10^{-7}$), as well as, the norm of the Cartesian gradient ($10^{-4}$ au). All of the DFT calculations were performed using the Gaussian package. The SCS(MI)-MP2 calculations were performed using Turbomole program package.

## Acknowledgments

This research has been financed by Greek National funding through the Operational Program ''Education and Lifelong Learning'' of the National Strategic Reference Framework (NSRF) - Research Funding Program: Thales ("Graphene and its Nanocomposites: Production, Properties and Applications" – Grant no: MIS 380389). The financial support of the European Research Council through the projects ERC AdG 2013 ("Tailor Graphene" ) and ERC StG 2011 (BIHSNAM) and ERC PoC 2013 (REPLICA2) is also acknowledged. Otakar Frank acknowledges a Czech Science Foundation project (nr. 14-15357S). Georgia Tsoukleri acknowledges the financial support of Heracleitus-II PhD grant. Finally, all authors acknowledge the financial support of the Graphene FET Flagship ("Graphene-Based Revolutions in ICT And Beyond"- Grant agreement no: 604391).

## Author contribution statement

Project planning, C.G.; sample preparation Ch.A., K.S.N.; spectroscopic measurements and analysis, Ch.A., O.F., G.T., J.P., K.P.; data interpretation and modelling Ch.A., D.S., E.N.K., N.P., C.G.; manuscript drafting, C.G., Ch.A., E.N.K., D.S.

## Additional information.

Supplementary Information accompanies this paper at http://www.nature.com/scientificreports.

Competing financial interests: The authors declare no competing financial interests.

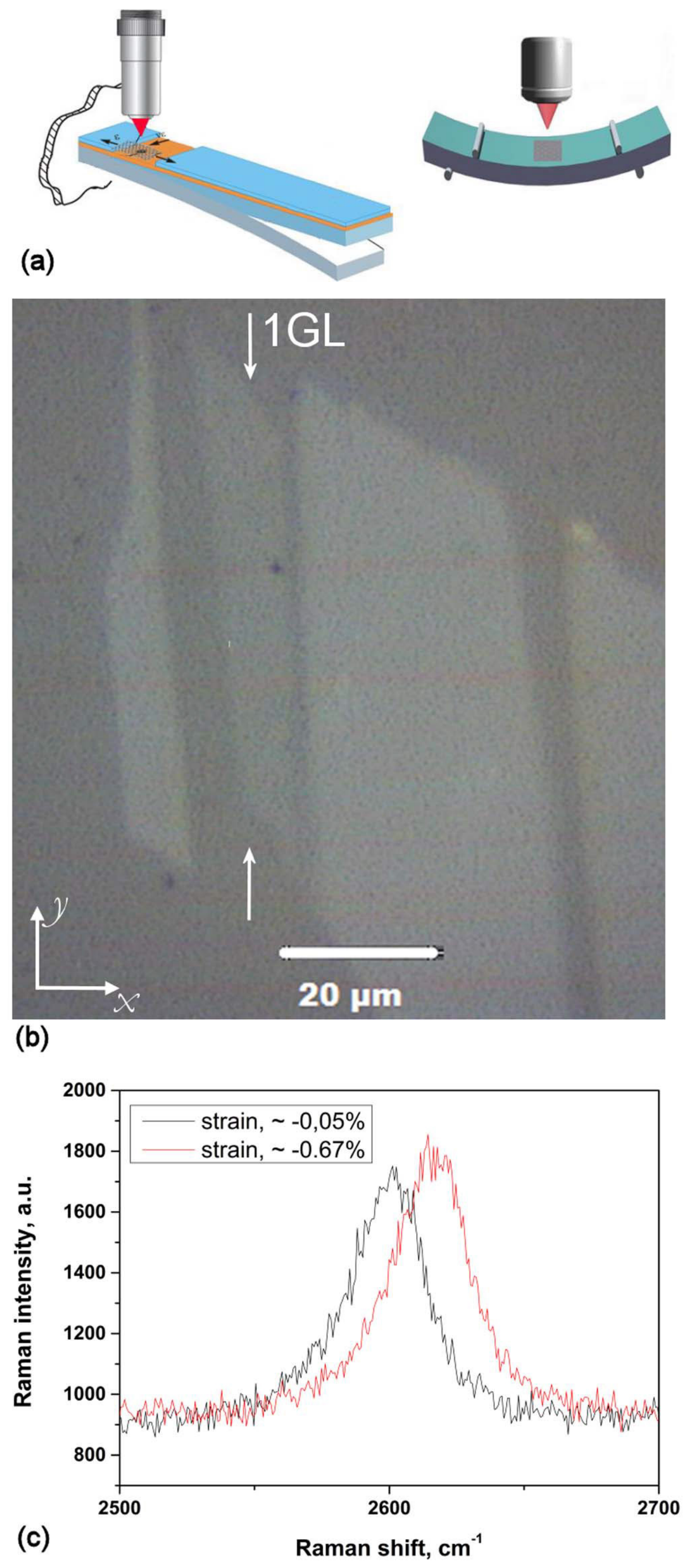

**Figure 1**.(a) Experimental configurations of cantilever beam and 4 point bending setup employed for the execution of the experiments. (b) Image taken with an optical microscope; the flake dimensions were $l$= 11μm and $w$=50μm (c) representative Raman spectra of the 2D peak measured at the onset of the experiment and just prior to failure. The shift to higher wave number is clearly seen.

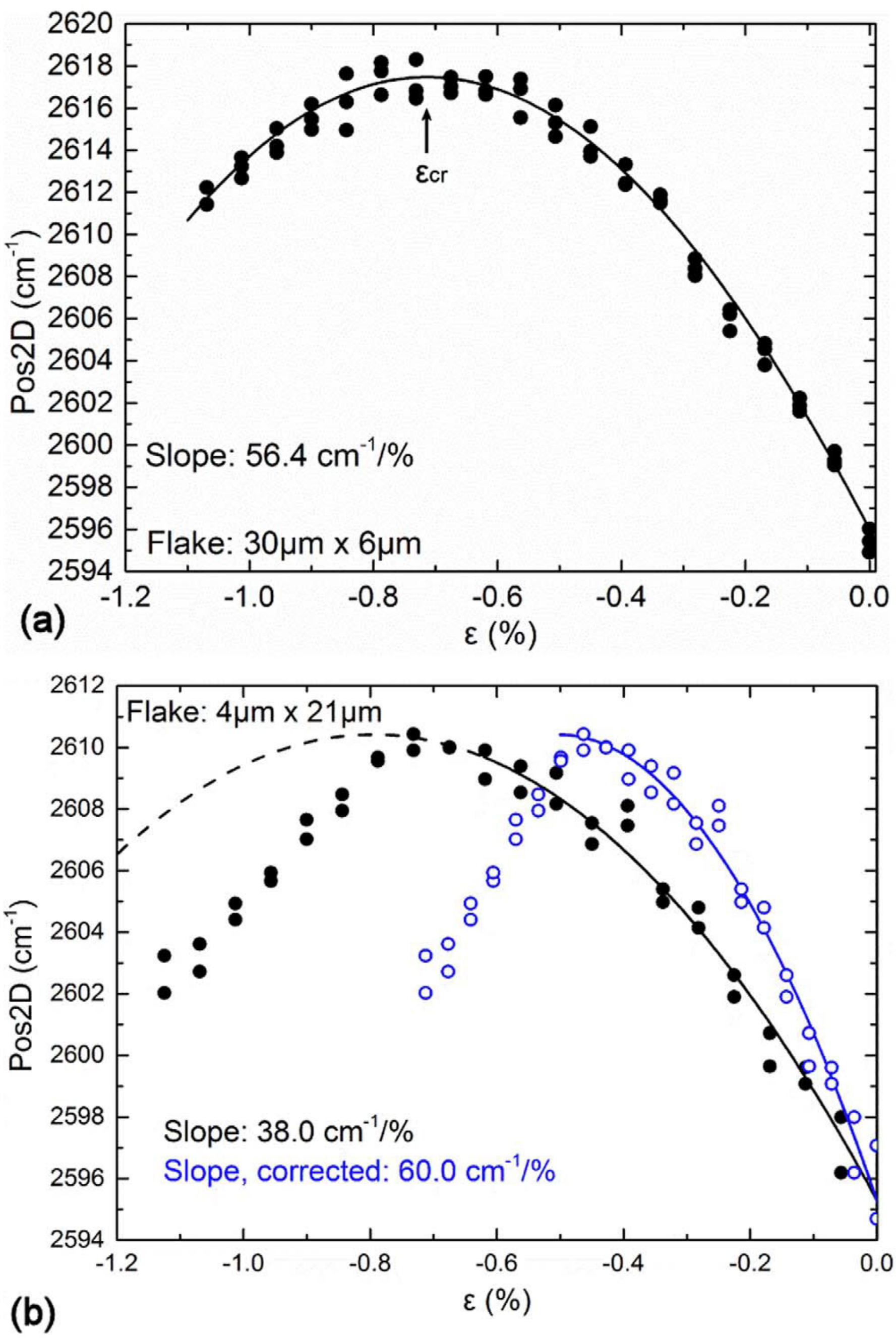

**Figure 2.**(a)Typical curve of Raman wavenumber of the 2D peak versus applied strain for a flake with $l$=30μm and $w$=6 μm.The line is a fourth order polynomial and fitting to the datais given by $Pos2D$=2596 – 56.41$\varepsilon$– 29.54$\varepsilon^2$ + 13.16$\varepsilon^3$ + 4.01$\varepsilon^4$. The plateau is clearly identified and the critical strain is calculated at 0.7%. (b) dependence of the 2D Raman peak on the strain for the flake with $l$=4μm and $w$=21μm. The open circles correspond to the applied strain of the beam and the solid circles represent the corrected strain values in accordance to equation 1. The black line is described by the Equation $Pos2D_{black}$=2595.3 – 38.0$\varepsilon$ – 23.9$\varepsilon^2$, and the blue line is a rescaling to a slope of 60 $cm^{-1}$/%.

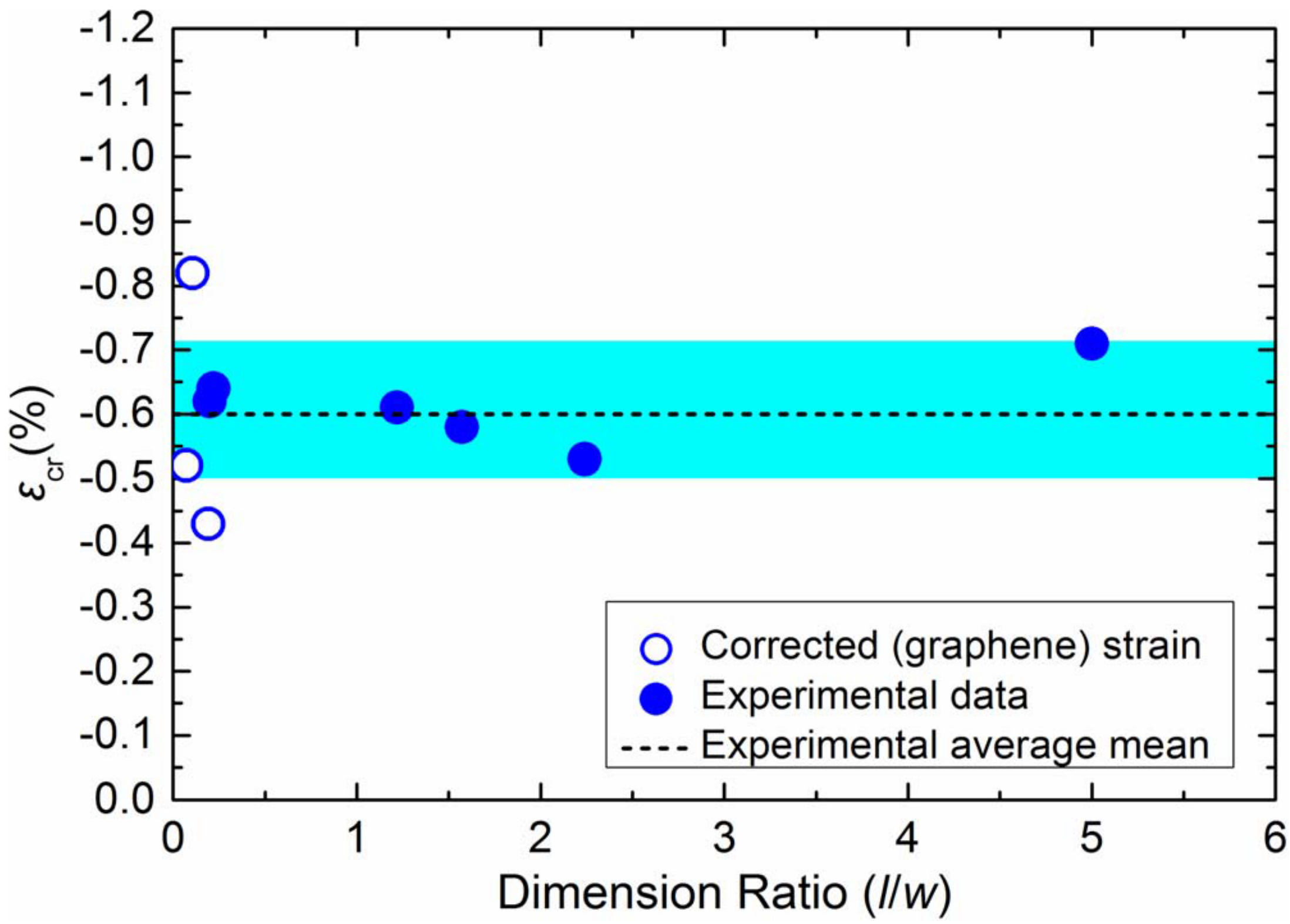

**Figure 3.**The critical strain for buckling is plotted versus the dimension's ratio *l/w*, for the experimental and values obtained from the Euler/Winkler theory. The experimental mean value is shown with the dash line and corresponds to a Winkler modulus of 6.7 GPa/nm. The shadowed area corresponds to the zone that describes the standard deviation from the mean value (±0.11 %).

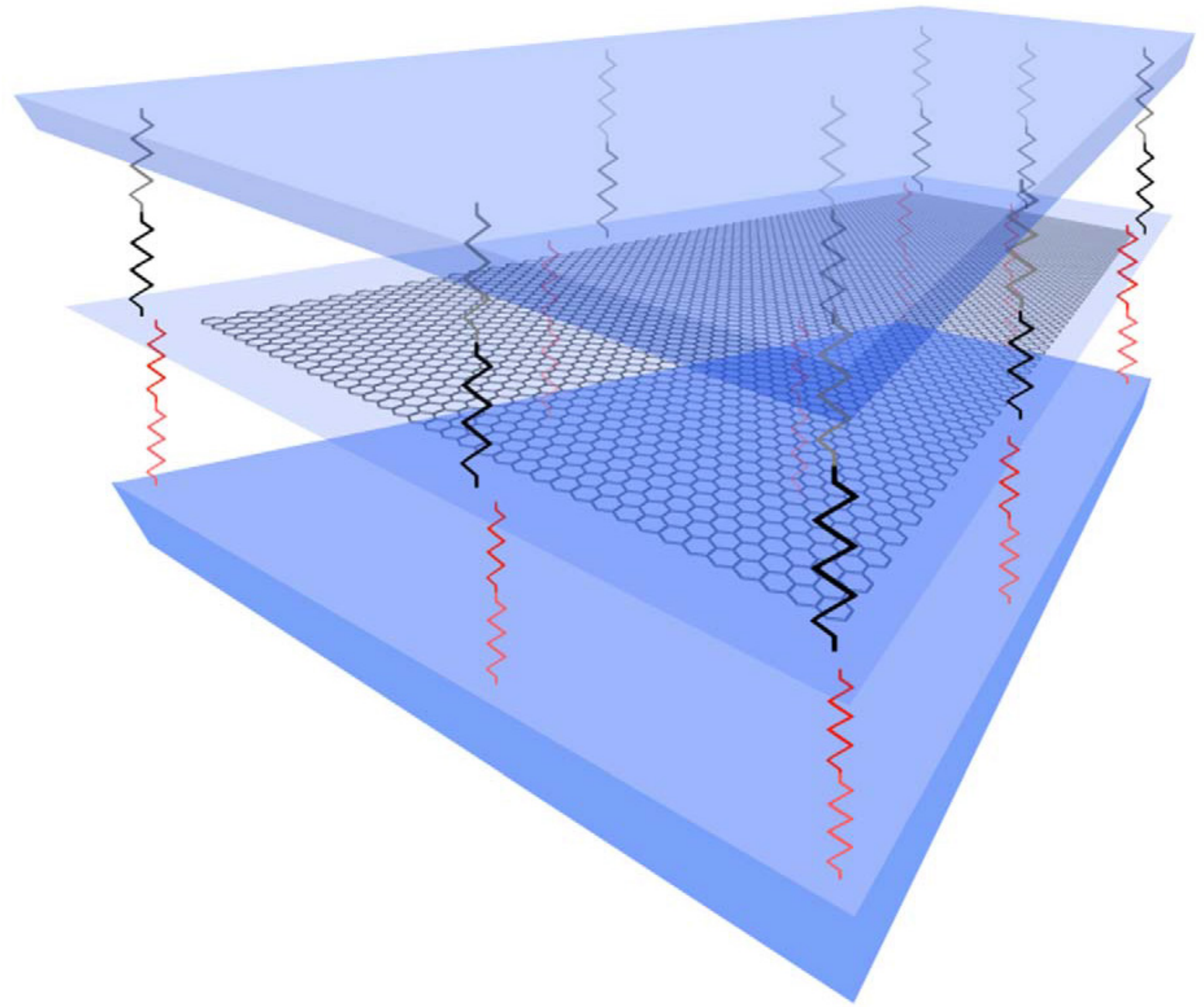

**Figure 4.** The interaction is modeled with linear springs with modulus $K_w$. In the case of the embedded graphene both sides of the flake are in contact with the polymer matrix.

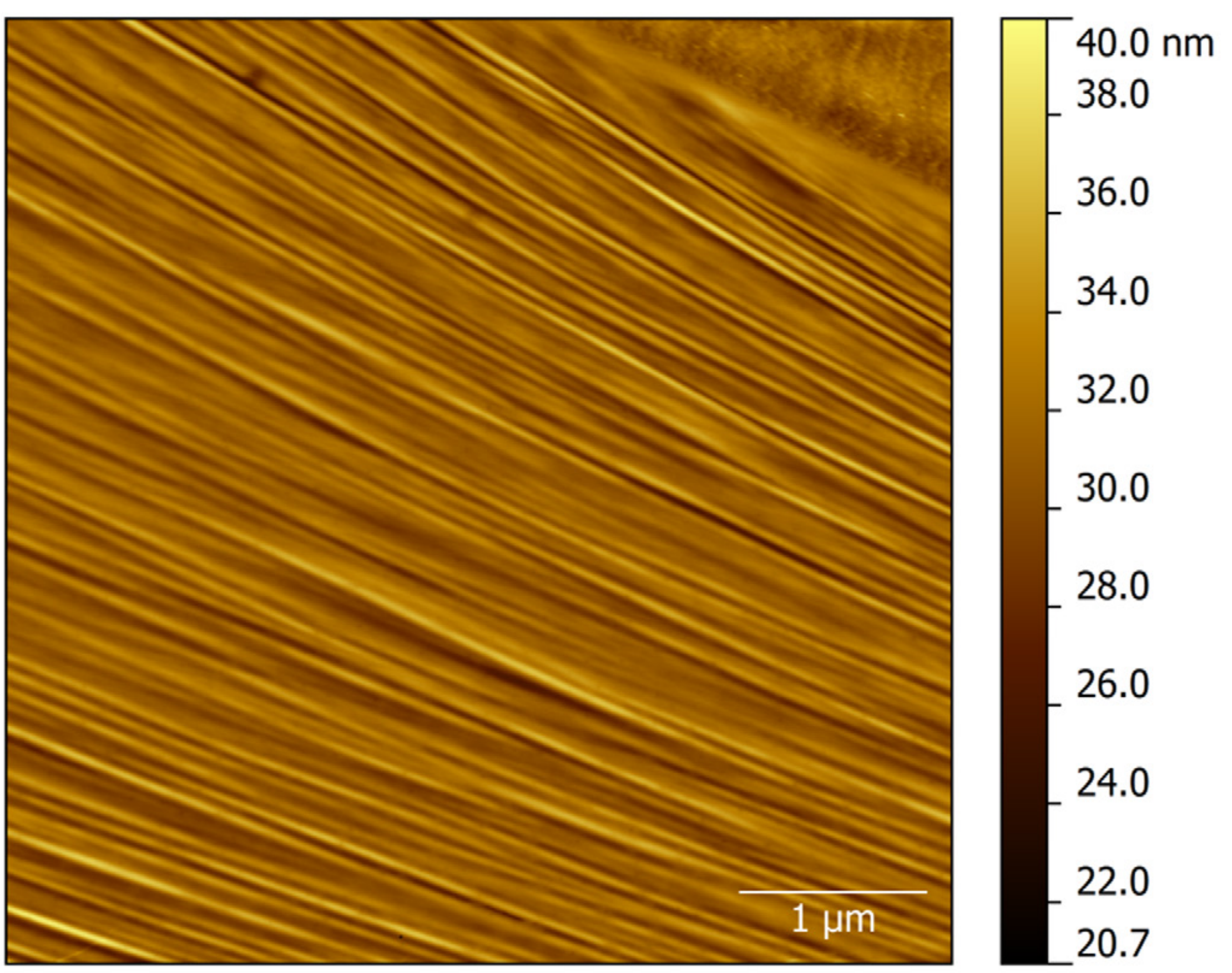

**Figure 5.** Wrinkling (buckling) observed for simply supported monolayer graphene flake under compression. In this case the flake is resting on SU8 photoresist polymer. The average half-wavelength in this case is about $\lambda$=50 nm at a height of about 2 nm.

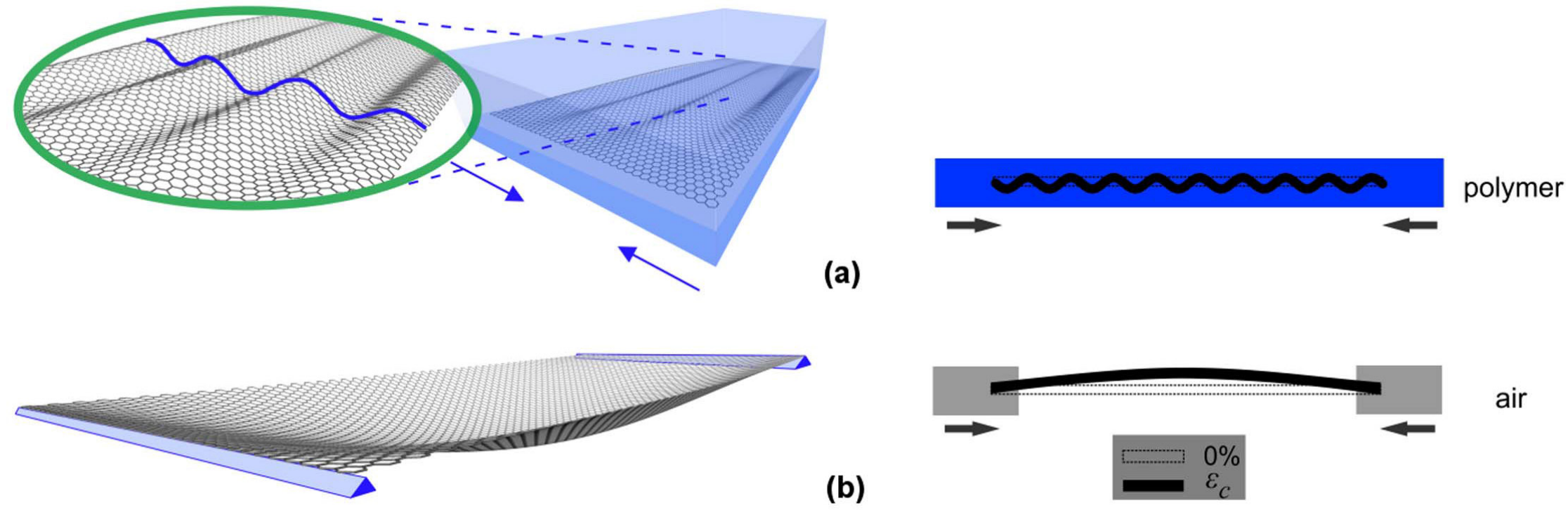

**Figure 6.**(a) Embedded graphene flake under compression that fails with multiple rippling with small wavelength, (b) Flake buckling in air for *m=1*.For clarity the rippling amplitudes are not under scale.

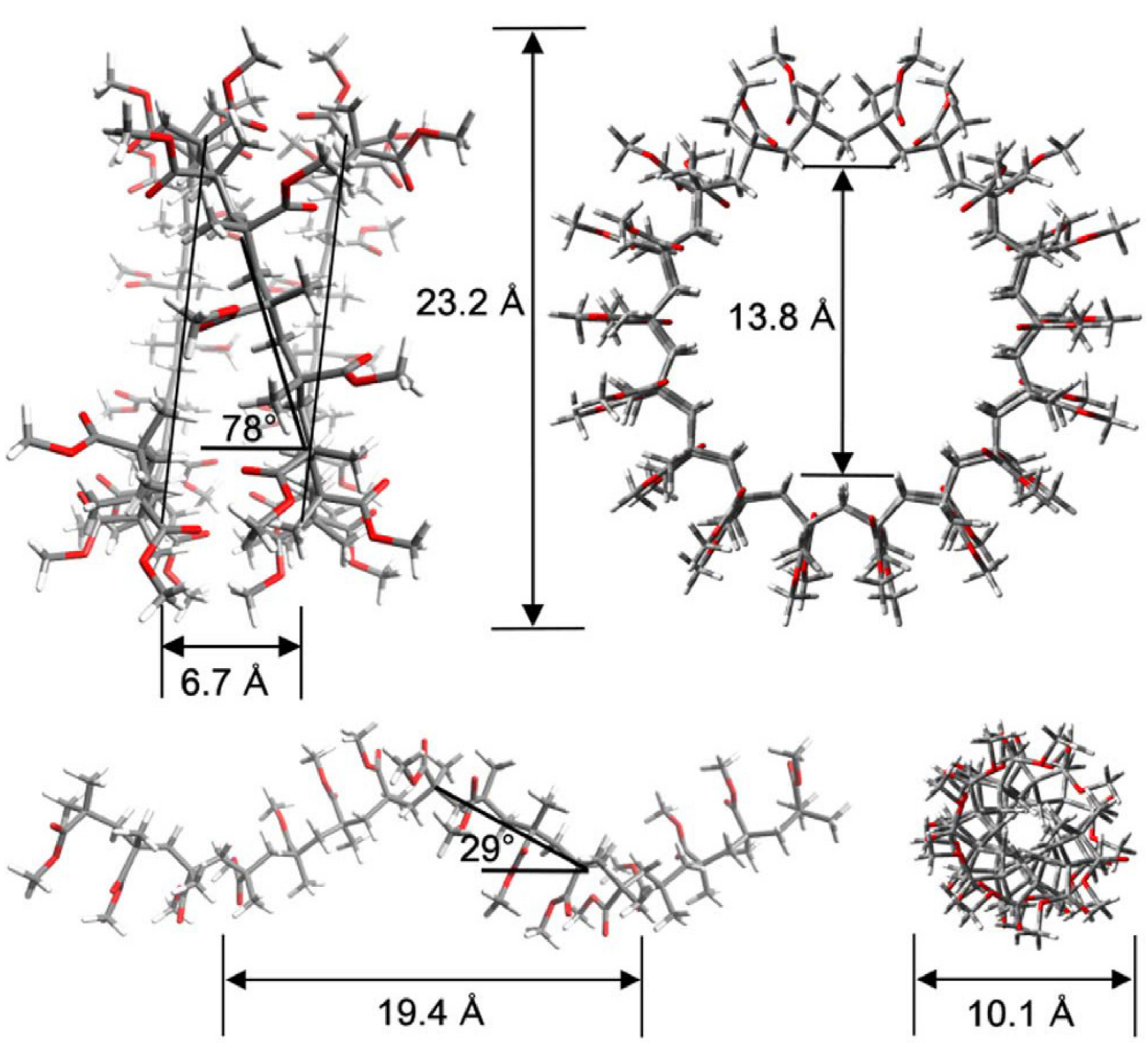

**Figure 7.** Axial and face view of a (top) $32_l$ s-PMMA and a (bottom) $16_l$i-PMMA polymer chain. The structures are optimized at the DFT/B97-D/def-SVP level.

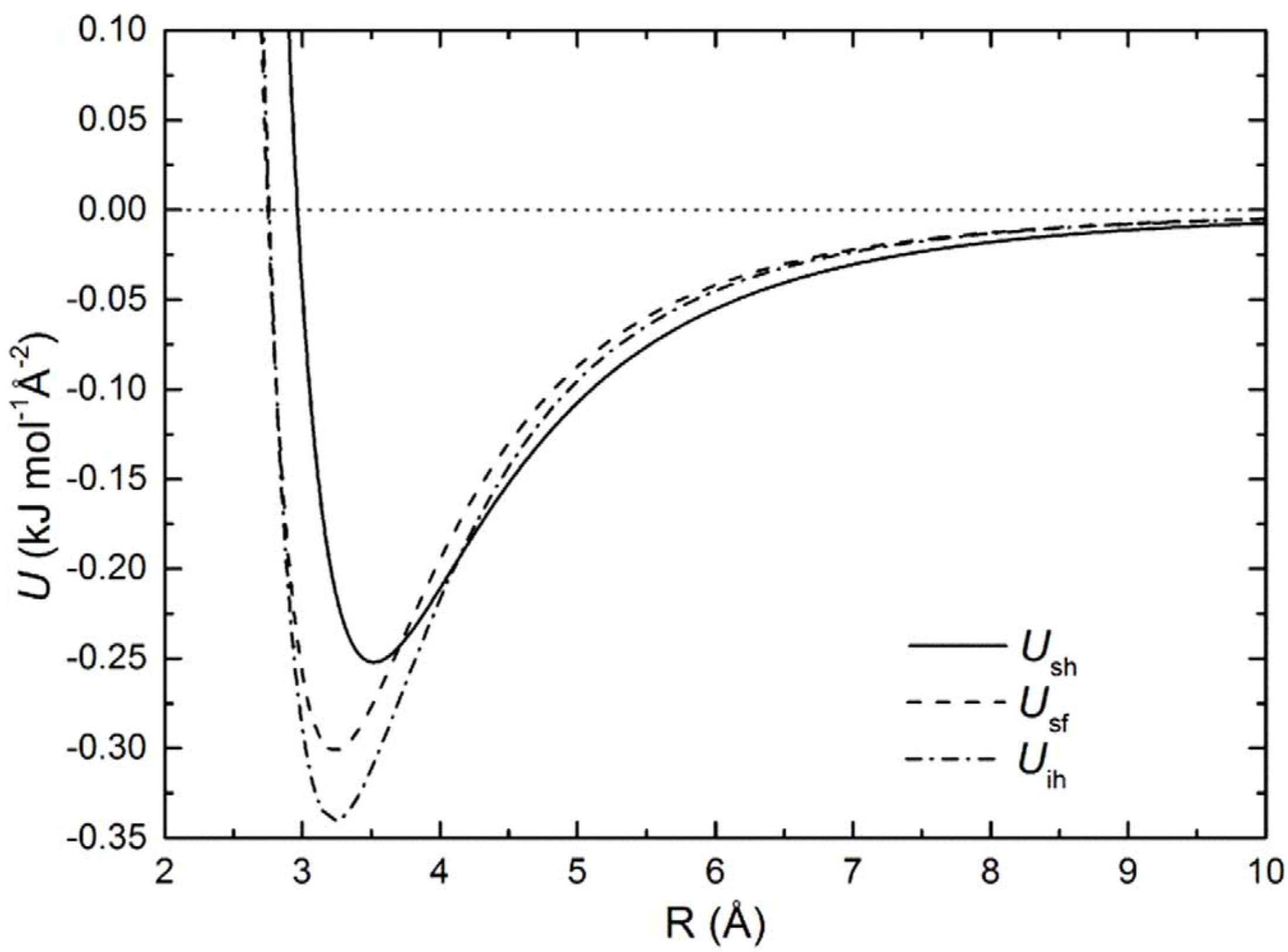

**Figure 8.**The composite interaction potentials per unit area. $U_{sh}$, for s-PMMA deposited horizontally on graphene, $U_{sf}$, for s-PMMA deposited face-down on graphene, and $U_{ih}$, for i-PMMA deposited horizontally on graphene. Dense depositions are assumed.

**Table 1.** Full presentation of the critical strain for buckling and the geometry of every specimen examined here and previously (ref. 3). Strain correction has only been implemented for data the slope of which lies outside the boundaries of the standard deviation value of $\pm 5$ $cm^{-1}$/% from the mean absolute value (60 $cm^{-1}$/%).

| [a]Nominal applied strain at failure (%) | *l* (μm) | *w* (μm) | Configuration | $\|cm^{-1}/\%\|$ | Critical (graphene) Strain *(%)* | $K_w$ (GPa/nm) | Half-wave number, *m* | Half-wave length, *λ*(nm) |
|---|---|---|---|---|---|---|---|---|
| -0.67 | 5 | 70 | 4pb | 46.8 | **-0.52**[c] | 4.88 | 3742 | 1.33 |
| -1.25[b] | 6 | 56 | Cantilever | 39.4 | **-0.82**[c] | 12.14 | 5639 | 1.06 |
| -0.62 | 6 | 30 | 4pb | 60.1 | **-0.62** | 6.94 | 4742 | 1.26 |
| -0.68 | 4 | 21 | 4pb | 38.0 | **-0.45**[c] | 3.34 | 2722 | 1.46 |
| -0.64[b] | 11 | 50 | Cantilever | 55.1 | **-0.64** | 5.46 | 8467 | 1.29 |
| -0.61 | 28 | 23 | Cantilever | 69.6 | **-0.61** | 6.72 | 22701 | 1.22 |
| -0.53[b] | 56 | 25 | Cantilever | 59.1 | **-0.53** | 5.07 | 42314 | 1.31 |
| -0.71 | 30 | 6 | 4pb | 56.4 | **-0.71** | 9.36 | 26423 | 1.12 |
| -0.58 | 22 | 14 | 4pb | 60.3 | **-0.58** | 6.07 | 17388 | 1.26 |

a. Applied strain calculated from beam Equation. b. Data from reference 14.c. Corrected data for short transfer length (equation 1).

**Table 2.** Fitting parameters, *ε*, *σ*, and *C* of the modified Lennard–Jones composite potential, $U_{sh}$, $U_{sf}$ and $U_{ih}$, as well as the corresponding non-linear spring stiffness per unit area, *K*.

| Configuration | *ε* ($kJmol^{-1}Å^{-2}$) | *σ* (Å) | *C* | *K* ($kJmol^{-1}Å^{-4}$) | *K* (GPa/ nm) |
|---|---|---|---|---|---|
| *s*-PMMA horizontal | 0.25 | 2.96 | 0.67 | 0.66 | 11.0 |
| *s*-PMMA face-down | 0.30 | 2.75 | 0.71 | 1.04 | 17.3 |
| *i*-PMMA horizontal | 0.34 | 2.76 | 0.72 | 1.22 | 20.3 |